# Synthetic antiferromagnet with Heusler alloy Co$_2$FeAl ferromagnetic layers

X. G. Xu, D. L. Zhang, X. Q. Li, J. Bao, Y. Jiang[*]

State Key Laboratory for Advanced Metals and Materials, School of Materials Science and Engineering, University of Science and Technology Beijing, Beijing 100083, China.

*PACS*: 75.70.-i; 75.70.Cn; 75.60-d; 75.10-b

Abstract

Heusler alloy Co$_2$FeAl was employed as ferromagnetic layers in Co$_2$FeAl/Ru/Co$_2$FeAl synthetic antiferromagnet structures. The experimental results show that the structure with a Ru thickness of 0.45 nm takes on strongly antiferromagnetic coupling, which maintains up to 150 $^o$C annealing for 1 hour. The structure has a very low saturation magnetization M$_s$ of 425 emu/cc, a low switching field H$_{sw}$ of 4.3 Oe and a high saturation field H$_s$ of 5257 Oe at room temperature, which are favorable for application in ultrahigh density magnetic read heads or other magnetic memory devices. XRD study testifies that the as-deposited Co$_2$FeAl film is in *B*2 phase. Therefore Heusler alloys can be used to fabricate SyAF and it is possible to make "all-Heusler" spin-valves or magnetic tunneling junctions with better magnetic switching properties and high magnetoresistance.

---

[*] To whom correspondence should be addressed. Email: yjiang@mater.ustb.edu.cn

The development of nanometer scale spintronics devices such as magnetic random access memories (MRAMs) and magnetoresistive read heads is one of the key points deciding the next generation magnetic storage density. However, with the rapid shrinking of device size, the switching field ($H_{sw}$) increases dramatically due to the demagnetizing field from the cell edge.[1] Synthetic antiferromagnet (SyAF) provides a solution to this problem. SyAF is a trilayer film consisting two ferromagnetic (FM) layers separated by an ultrathin Ru layer less than 1 nm, in which two FM layers couple antiferromagnetic strongly.[2] Thus, SyAF structure takes on a closed magnetic flux configuration and a low stray field, which can reduce magnetostatic coupling between SyAF and neighboring FM layers.[3] Because of these priorities, SyAF can be applied in spin valves (SPVs) or magnetic tunneling junctions (MTJs) as free layers to obtain so-called size-independent $H_{sw}$.[4] In recent years, researchers have pulled out continuous efforts to optimize properties of SyAF.[5,6] The most popular and well studied SyAF structure CoFe/Ru/CoFe has excellent antiferromagnetic coupling property and large saturation field ($H_s$) above 10000 Oe. The large saturation magnetization ($M_s$) and coercivity ($H_c$) of CoFe result in a high $H_{sw}$ value about 60 Oe of SyAF CoFe/Ru/CoFe, which is a handicap for its application in high density MRAMs, as well as in magnetoresistive read heads. To lower the $H_{sw}$ value of SyAF structure, it is necessary to choose a FM layer with low $H_c$ and $M_s$ values, such as NiFe alloy which has a typical $H_c$ lower than 1 Oe. However, the low $H_{sw}$ value of SyAF NiFe/Ru/NiFe is obtained on the cost of poor antiferromagnetic coupling.[7] Lee et al.[8] reported a NiFe-based SyAF structure in which an ultrathin CoFe layer of 0.2 nm was inserted between NiFe and Ru layers. The MTJ with a free layer of composite SyAF containing NiFe and CoFe has realized a relative low $H_{sw}$ value. However, the five layers composite SyAF structure needs a complicated fabrication process.

Considering the antiferromagnetic coupling origination of SyAF from Co/Ru/Co trilayer[2] and the high spin polarization of Heusler alloy $Co_2FeAl$,[9] it is possible to employ Co-rich $Co_2FeAl$ in SyAF as FM layers. This kind of 'Heusler synthetic antiferromagnet' (HSyAF) should strongly improve magnetoresistance when it is applied in SPVs or MTJs. Ishikawa et al.[10] studied the exchange bias effect in $Co_2Cr_{0.6}Fe_{0.4}Al$/Ru/$Co_{90}Fe_{10}$/IrMn layer structure, which also shows SyAF property. However, its $H_s$ value about 1000 Oe is too weak for practical application. In this paper, we report our experimental works on HSyAF, which is demonstrated to have not only strong antiferromagnetic coupling between two $Co_2FeAl$ layers, but also low $H_{sw}$ and $M_s$ values.

Since the property of $Co_2FeAl$ is sensitive to the crystal structure of film, we deposited $Co_2FeAl$ (30) (the numbers in parentheses are the nominal thickness in nanometers) on silicon (100) with Ta (4) as buffer and cap layers at room temperature (RT) by magnetron sputtering in a uniform dc field. The base pressure before deposition was lower than $2.0 \times 10^{-5}$ Pa and the argon pressure was kept at $5 \times 10^{-1}$ Pa during sputtering. The crystal structure of as-deposited films was characterized by X-ray diffraction (XRD) to ensure that $Co_2FeAl$ component in HSyAF maintains Heusler alloy crystal structure. HSyAF samples Ta (4)/$Co_2FeAl$ (3)/Ru (*x*)/$Co_2FeAl$ (5)/Ta (4) (*x*=0.45, 0.65, 1) were deposited on silicon (100) substrate under the same experimental condition as that of the $Co_2FeAl$ depositing. The as-deposited samples were annealed at temperatures of 100, 150, 200, and 250 °C for 1 h respectively in vacuum furnace with a pressure lower than $5 \times 10^{-5}$ Pa. For comparison, we also fabricated several traditional SyAF samples with structures of Ta (4)/$Co_{90}Fe_{10}$ (3)/Ru (0.45)/$Co_{90}Fe_{10}$ (5)/Ta (4), Ta (4)/NiFe (3)/Ru (0.45)/NiFe (5)/Ta (4), and Ta (4)/NiFe (2.5)/$Co_{90}Fe_{10}$ (0.5)/Ru (0.45)/$Co_{90}Fe_{10}$ (0.5)/NiFe (4.5)/Ta (4), with different annealing temperatures of 100, 200, and 300 °C. The magnetic properties of both as-deposited and annealed samples with a size of 3×3 mm$^2$ were measured by alternating gradient magnetometer (AGM) at RT.

The XRD spectrum of as-deposited Ta (4)/Co$_2$FeAl (30)/Ta (4) at RT is presented in Figure 1. The peak locating at 70$^o$ corresponds to the Si substrate, and the peak around 38$^o$ belongs to the buffer and cap layers Ta. The diffraction pattern shows two Co$_2$FeAl peaks (200) and (220), while the (220) peak is much weak relative to the (200) peak, which indicates that the Co$_2$FeAl film exhibits highly directional growth in (200). XRD pattern of Heusler alloy Co$_2$FeAl $L2_1$ phase should have two superlattice reflection peaks (111) and (200) because of highly ordered Fe and Al atoms in addition to fundamental reflection peaks such as (220) peak. In $B2$ structure, the atom disorder of Fe and Al causes the intensity reduction of odd superlattice reflections (111). While the even peak (200) will decrease when part of Co sites are substituted by Fe or Al atom in $A2$ structure. So the (111) and (200) peaks represent $L2_1$ and $B2$ structure, respectively.[11] Accordingly, the highly directional as-deposited Co$_2$FeAl film presents $B2$ structure.

The magnetization curve of as-deposited Co$_2$FeAl (30) sample is presented in Figure 2. The $M_s$ value of Co$_2$FeAl thin film about 1000 emu/cc is close to that of bulk phase, which demonstrates the good quality of the as-deposited Co$_2$FeAl film. The magnetization curves of HSyAF Ta (4)/Co$_2$FeAl (3)/Ru ($x$)/Co$_2$FeAl (5)/Ta (4) ($x$=0.45, 0.65, 1) are shown in Figure 3, while the inset is the corresponding $M$-$H$ minor loops. It is obviously that the Ta (4)/Co$_2$FeAl (3)/Ru (0.45)/Co$_2$FeAl (5)/Ta (4) sample takes on strongly antiferromagnetic coupling with a saturation field H$_s$ of 5257 Oe and a very low saturation magnetization M$_s$ of 425 emu/cc. The M$_s$ is much lower than that of the single Co$_2$FeAl film (Fig. 2), which is very similar to the behavior we observed in Co$_{90}$Fe$_{10}$ based SyAF.[12] One reason is that the effective thicknesses of the $Co_2FeAl$ films in the SyAF are much smaller than their nominal values because of the intermixing of the $Co_2FeAl$ and Ru layers. Another reason is possibly a canting of magnetic moments near the $Co_2FeAl$/Ru interfaces because of the strong antiferromagnetic coupling between neighboring $Co_2FeAl$ layers. The canting does not completely disappear even when the SyAF is put in a very strong magnetic field. When the Ru thickness $x$ increases to 0.65 nm, the $H_s$ value of HSyAF decreases to 750 Oe, which indicates that the antiferromagnetic coupling strength is much weaker than that of $x$ = 0.45 nm sample. The antiferromagnetic coupling almost disappears when the Ru thickness $x$ increases to 1 nm, of which the *M-H* curve shows FM feature. Therefore, the antiferromagnetic coupling strength is sensitive to Ru thickness, and a stronger antiferromagnetic coupling depends on a thinner Ru layer. However, the Ru layer of 0.45 nm contains only 3 atomic layers in average, and when the Ru layer becomes thinner, it will be broken and lose antiferromagnetic coupling configuration, especially after annealing. So, the HSyAF with Ru thickness of 0.45 nm possesses the most favorable antiferromagnetic coupling property, which is similar to that of the traditional SyAF. From the inset of Figure 3, the $H_{sw}$ values are 4.3, 4.6, and 1.4 Oe, when the values of Ru thickness $x$ are 0.45, 0.65, and 1 nm respectively, which are low enough for magnetoresistive read heads.

To investigate thermal stability of HSyAF, Ta (4)/$Co_2FeAl$ (3)/Ru (0.45)/$Co_2FeAl$ (5)/Ta (4) sample was annealed at different temperatures for 1 hour, and its *M-H* curves are measured at RT and presented in Figure 4. Comparing the four *M-H* curves, we found that the antiferromagnetic coupling strength decreases with the annealing temperature increasing. The two $Co_2FeAl$ layers keep antiferromagnetic coupling after 1 hour annealing at 100 °C, with only slightly $H_s$ decrease of 300 Oe relative to the as-deposited sample. When the annealing temperature increases to 150 °C, the $H_s$ value is still 2889 Oe. However, after 200 °C annealing for 1 hour, the sample only has a small $H_s$ value of 549 Oe and shows poor antiferromagnetic coupling feature. The sample annealed at 250 °C for 1 hour (The curve is not shown in Figure 4) shows the typical ferromagnetic *M-H* curve feature. So, the HSyAF sample can keep a strong antiferromagnetic coupling after being annealed up to 150 °C for 1 hour. Along with the increasing annealing temperature, the saturation magnetization $M_s$ increases in company with the reducing of $H_s$ from 425 Oe to 623 Oe. The changes of $M_s$ with annealing temperature are related with the antiferromagnetic coupling strength. When the HSyAF sample is annealed, antiferromagnetic coupling strength becomes weaker. The canting moments are therefore less than those in the as-deposited HSyAF.

Traditional SyAF samples with $Co_{90}Fe_{10}$, NiFe, and $Co_{90}Fe_{10}$/NiFe as FM layers are also fabricated at the same experimental conditions to compare with the HSyAF samples. The saturation field $H_s$ as a function of the annealing temperature is presented in Figure 5. The $H_s$ value of the HSyAF is lower than those of the SyAF made of $Co_{90}Fe_{10}$ and $Co_{90}Fe_{10}$/NiFe but higher than that of the one made of NiFe at each annealing temperature. However, the antiferromagnetic coupling strength of Ta (4)/$Co_2FeAl$ (3)/Ru (0.45)/$Co_2FeAl$ (5)/Ta (4) is good enough for application. The saturation magnetization $M_s$ of the HSyAF and SyAF samples is shown via annealing temperature in Figure 6. It is obviously, that the SyAF with $Co_{90}Fe_{10}$ FM layers has large $M_s$ values at the whole annealing temperature region. While the $M_s$ of Ta (4)/$Co_2FeAl$ (3)/Ru (0.45)/$Co_2FeAl$ (5)/Ta (4) is the lowest at RT and comparable with that of the one-hour-annealed SyAF samples with NiFe and $Co_{90}Fe_{10}$/NiFe as FM layers. The very low $M_s$ of HSyAF is helpful to reduce $H_{sw}$, especially when the cell size is reduced down to nanometer, so as to satisfy the requirement of ultrahigh storage density.

In summary, we have successfully fabricated the HSyAF structures of $Co_2FeAl/Ru/Co_2FeAl$ and demonstrated a strong antiferromagnetic coupling between the two $Co_2FeAl$ layers when the Ru layer thickness is 0.45 nm. The HSyAF with the structure of Ta (4)/$Co_2FeAl$ (3)/Ru (0.45)/$Co_2FeAl$ (5)/Ta (4) has an $H_s$ of 5257 Oe at RT, and the thermal stability study indicates that the strong antiferromagnetic coupling can maintain up to 150 $^{o}$C. Moreover, the HSyAF has very low $M_s$ of 425 emu/cc and $H_{sw}$ of 4.3 Oe, which are favorable for application in ultrahigh density magnetoresistive read heads or other magnetic memory devices. The crystal structure study indicates that the as-deposited $Co_2FeAl$ is in Heusler *B*2 phase. These results demonstrate a fact that Heusler alloys can also be used to fabricate SyAF and it is possible to make "all-Heusler" spin-valves or magnetic tunneling junctions with high magnetoresistance if we can realize *A*2 $Co_2FeAl$ structure or substituting part of Al with Si.

This work was supported by the NSFC (Grant Nos. 50701005, 50831002), the Keygrant Project of Chinese Ministry of Education.(No. 309006) and the Ph.D. Programs Foundation (Grant No. 20070008024) of the Chinese Ministry of Education, the National Basic Research Program of China (Grant No. 2007CB936202).

Figure Captions

FIG. 1. XRD spectrum of as-deposited $Co_2FeAl$ (30 nm) on silicon (100) with Ta (4 nm) as buffer and cap layers.

FIG. 2. *M-H* loop for the as-deposited $Co_2FeAl$ (30 nm) on silicon (100) with Ta (4 nm) as buffer and cap layers.

FIG. 3. (Color online) *M-H* loops for the HSyAF of Ta (4)/$Co_2FeAl$ (3)/Ru (*x*)/$Co_2FeAl$ (5)/Ta (4) (*x*=0.45, 0.65, 1) (in nm) measured at RT. The inset shows the *M-H* curves for the minor loops.

FIG. 4. (Color online) *M-H* loops for as-deposited and annealed HSyAF samples Ta (4)/$Co_2FeAl$ (3)/Ru (0.45)/$Co_2FeAl$ (5)/Ta (4) measured at RT.

FIG. 5. (Color online) Saturation field $H_s$ of the SyAF of Ta (4)/FM (3)/Ru (0.45)/FM (5)/Ta (4) (in nm) as a function of annealing temperature. The FM layers are $Co_2FeAl$ (▼), NiFe (■), $Co_{90}Fe_{10}$ (●), or NiFe/$Co_{90}Fe_{10}$ (▲), respectively.

FIG. 6. (Color online) Saturation magnetization $M_s$ of the SyAF of Ta (4)/FM (3)/Ru (0.45)/FM (5)/Ta (4) (in nm) in terms of annealing temperature. The FM layers are $Co_2FeAl$ (▼), NiFe (■), $Co_{90}Fe_{10}$ (●), or NiFe/$Co_{90}Fe_{10}$ (▲), respectively.

Figure 1. (X. G. Xu et al.)

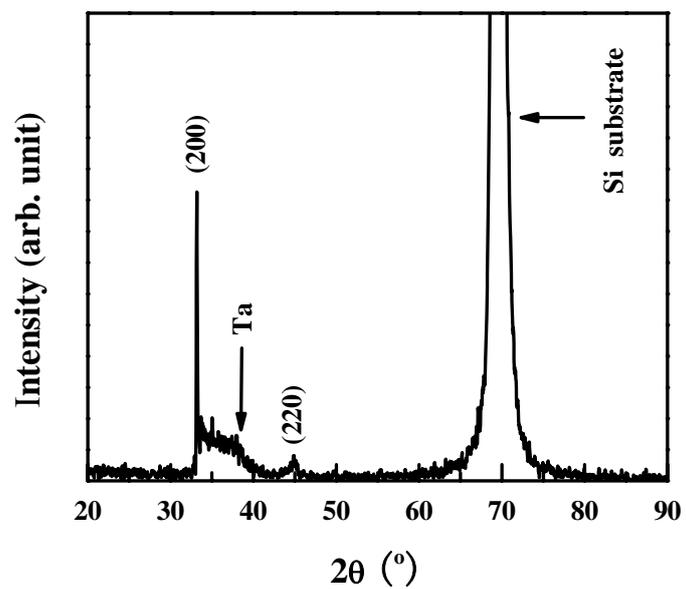

Figure 2. (X. G. Xu et al.)

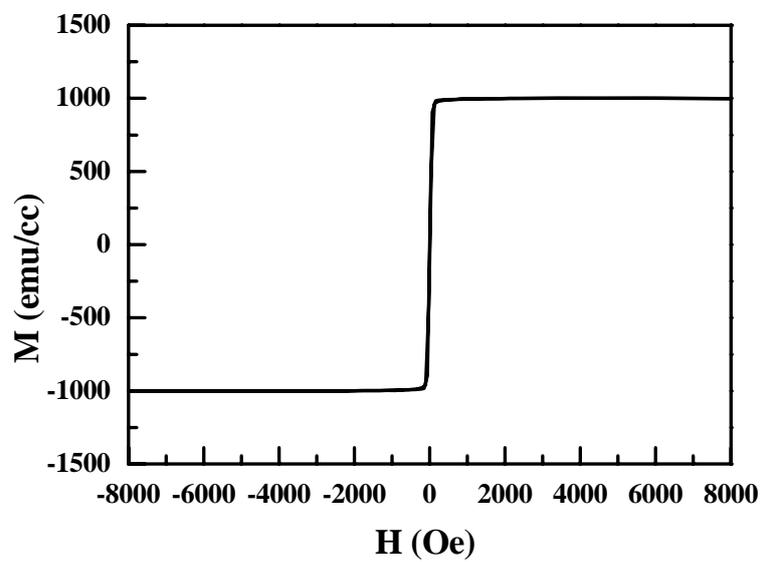

Figure 3. (X. G. Xu et al.)

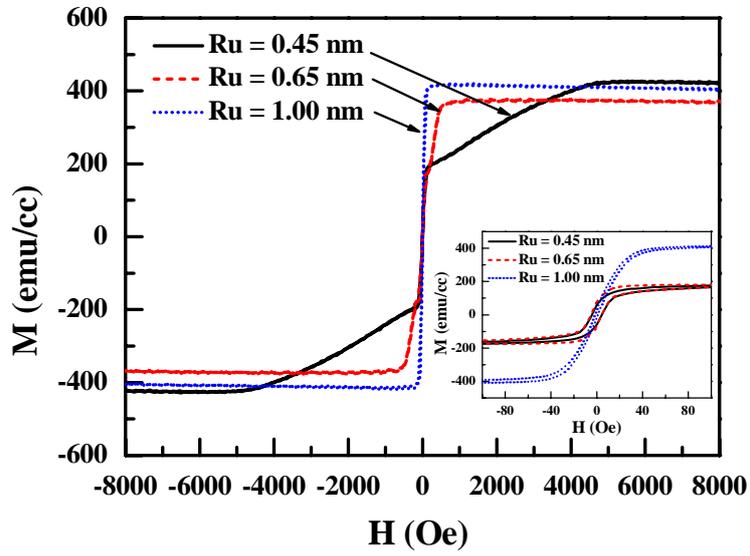

Figure 4. (X. G. Xu et al.)

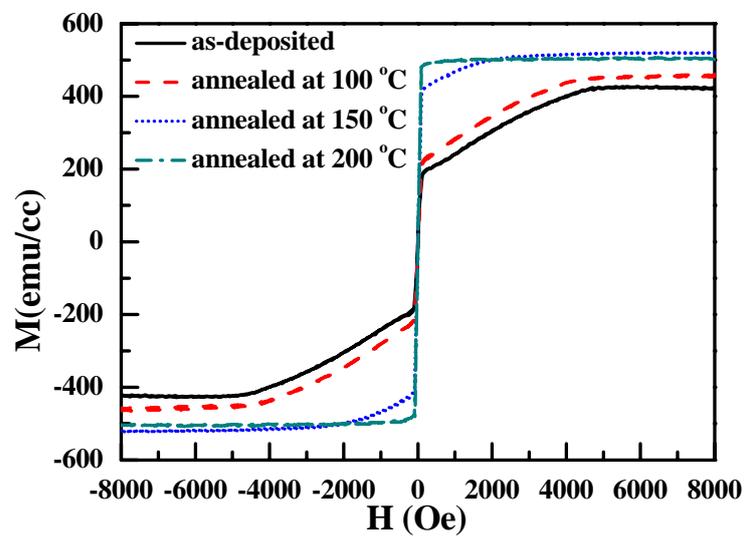

Figure 5. (X. G. Xu et al.)

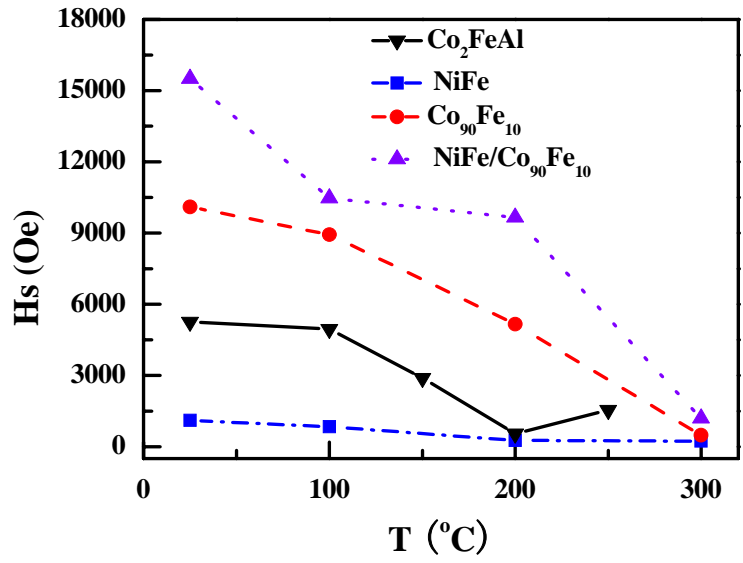

Figure 6. (X. G. Xu et al.)

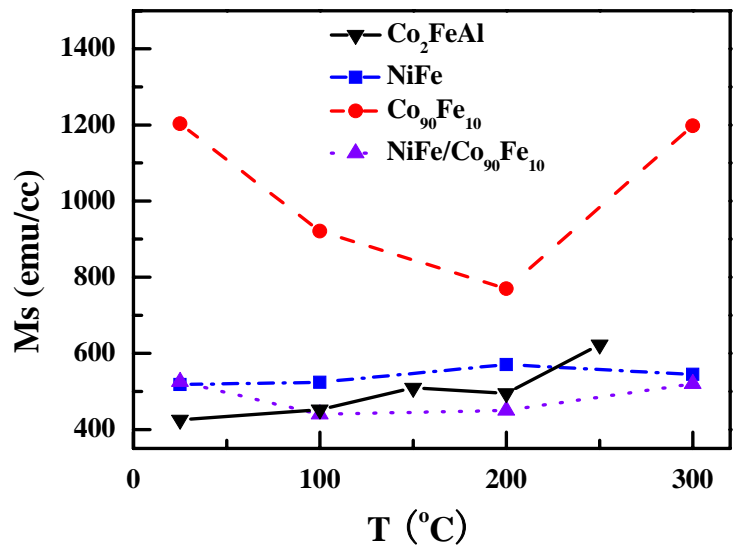